\documentstyle[onecolumn,epsfig,mnras_cite]{mn}
\title{Full-sky correlations of peaks in the microwave background}
\author[Alan F. Heavens and Sujata Gupta]
{Alan F. Heavens and Sujata Gupta\\
Institute for Astronomy, University of Edinburgh, Blackford Hill,
Edinburgh EH9 3HJ, U.K.}

\newcommand{\be}{\begin{equation}}
\newcommand{\ee}{\end{equation}}

\newcommand{\ba}{\begin{eqnarray}}
\newcommand{\ea}{\end{eqnarray}}

\newcommand{\nn}{\nonumber\\}

\newcommand{\bkk}[1]{{\bf{k}}_{ #1 }}

\newcommand{\bv}{{\bf v}}

\def\gs{\mathrel{\raise1.16pt\hbox{$>$}\kern-7.0pt 
\lower3.06pt\hbox{{$\scriptstyle \sim$}}}}         
\def\ls{\mathrel{\raise1.16pt\hbox{$<$}\kern-7.0pt 
\lower3.06pt\hbox{{$\scriptstyle \sim$}}}}         

\begin{document}

\maketitle

\begin{abstract}
We compute precise predictions for the two-point correlation function
of local maxima (or minima) in the temperature of the microwave
background, under the assumption that it is a random gaussian field.
For a given power spectrum and peak threshold there are no adjustable
parameters, and since this analysis does not make the small-angle
approximation of \scite{HS99}, it is essentially complete.  We find
oscillatory features which are absent in the temperature
autocorrelation function, and we also find that the small-angle
approximation to the peak-peak correlation function is accurate to
better than 0.01 on all scales.  These high-precision
predictions can form the basis of a sensitive test of the gaussian
hypothesis with upcoming all-sky microwave background experiments MAP
and Planck, affording a thorough test of the inflationary theory of
the early Universe.  To illustrate the effectiveness of the technique,
we apply it to simulated maps of the microwave sky arising from the
cosmic string model of structure formation, and compare with the
bispectrum as a non-gaussian discriminant.  We also show how peak
statistics can be a valuable tool in assessing and statistically
removing contamination of the map by foreground point sources.
\end{abstract}


\section{Introduction}

The cosmic microwave background radiation (CMB) presents an ideal
opportunity to test theories of the early Universe.  At the time of
last scattering, the Universe is a relatively straightforward, almost
uniform, mixture of photons, baryons, electrons and dark matter.  The
physics is well-understood, and free from the very complicated effects
which make interpretation of the present-day matter distribution more
complicated.  The microwave background thus offers the possibility of
accurately testing models of structure formation.  A generic test can
readily be made between two classes of structure-formation models,
based on inflation and cosmic defects respectively.  There are several
ways to do this; the power spectrum itself is a useful discriminant of
specific models.  We concentrate here on a generic test: most
inflationary models predict that the microwave background temperature
map will be very close to a random gaussian field, whereas generically
defect models predict a non-gaussian temperature map.  It turns out
that testing the gaussian nature of the initial fluctuations is easier
through analysis of CMB fluctuations than large-scale structure
\cite{VWHK00}, although tests based on number densities of
high-redshift objects may also be useful
(\pcite{Robinson00},\pcite{MVJ00}).

Current evidence from Boomerang \cite{Boomerang} and MAXIMA
(\pcite{MAXIMA1},\pcite{MAXIMA2}) favours
inflation models, since the power spectrum is acceptable for certain
combinations of cosmological parameters.  Indeed, the major scientific
goal of these and future experiments such as the {\em Microwave
Anisotropy Probe} (MAP) and {\em Planck Surveyor}
\cite{PhaseA}, is to derive cosmological parameters from the
power spectrum.  To make this interpretation requires that the
temperature map is created by inflation or some similar process, not
by defects, and that the map is not seriously contaminated by
foregrounds.  In both of these areas, the statistics of peaks can be a
valuable tool.  The process is quite straightforward: given a power
spectrum, the statistical properties of peaks of a gaussian field are
fully determined - there are no free parameters.  If the peaks are not
consistent with the predictions, then either the CMB temperature map is
not gaussian, or it is significantly contaminated by foregrounds, or
both.  In either of these cases, the derived cosmological parameters
from the power spectrum will be suspect.  In this paper, we compute
the predictions for the correlation function of local maxima (and
minima) for a gaussian field.  The paper generalises the work of \scite{HS99}
in dropping the small-angle approximation: the results of this paper
can be used for all valid separations on the sky.  There are several
ways to test the gaussian hypothesis, such as the three-point function
(e.g. \pcite{Hinshaw94}, \pcite{FRS93}, \pcite{LS93},
\pcite{GLMM94}), the genus and Euler-Poincar\'e statistic
(\pcite{Coles88}, \pcite{Gott90}, \pcite{Luo94b}, \pcite{Smoot94}),
the bispectrum (\pcite{Luo94}, \pcite{Hea98}, \pcite{Ferreira98}),
studies of tensor modes in the CMB \cite{CCT94}, excursion set
properties (\pcite{Barreiro98}, \pcite{Barreiro2000}), peak
statistics (\pcite{BE87}, \pcite{Kogut95}, \pcite{Kogut96},
\pcite{Barreiro97}) and wavelet analyses (e.g. \pcite{MHL00},
\pcite{AF99},
\pcite{FA99}).  One advantage which the method presented here
has is the possibility of assessing and removing contamination by
foreground point sources.  We return to this in the discussion.
Non-gaussian signals have been reported for the COBE map by
\scite{Ferreira98} (see also
\pcite{Pando98}, \pcite{KJ98}, \pcite{BT99}, \pcite{MHL00}, 
\pcite{Mag2000}) .  If this
nongaussian signal is really present in the microwave background map,
and not the result of some artefact \cite{Banday2000}, then it would
be a severe challenge to inflation models, as it is many orders of
magnitude larger than expected (e.g. \pcite{VWHK00} and references
therein).

\section{Method}

In this section, we compute the two-point correlation function of
local maxima in 2D gaussian random fields on the surface of a sphere.
The method essentially follows that of Heavens \& Sheth (1999), who
used a Fourier analysis which assumed a flat sky.  That analysis should be
accurate for small separations;  the analysis in this paper is general.

\subsection{Peaks on the surface of a sphere}

We define the temperature fluctuation by $\delta(\theta,\phi) \equiv
T(\theta,\phi)/\bar T -1$, where $\bar T$ is the mean temperature, and
its spherical harmonic transform by
\be
a_{\ell m} \equiv \int d^2\Omega\, \delta(\theta,\phi)
Y_\ell^{m*}(\theta,\phi)
\ee
where $\Omega=(\theta,\phi)$.  The inverse is
\be
\delta(\theta,\phi) = \sum_{m=-\ell,\ell;\,\ell=0,\infty} a_{\ell m}
Y_\ell^{m}(\theta,\phi)
\ee
If the temperature map is a random gaussian field, the statistical
properties of the fluctuations are specified entirely by the power
spectrum, $C_\ell$, defined by
\be
\langle a_{\ell m}a^*_{\ell'm'}\rangle = C_\ell
\delta^K_{\ell\ell'}\delta^K_{mm'}
\label{Cl}
\ee
where angle brackets indicate ensemble averages, and $\delta^K$ is a
Kronecker delta function.  The autocorrelation function of the
temperature for points at $(\theta,\phi)$, $(\theta',\phi')$,
separated by an angle $\psi$ is
\be
F(x) = \langle \delta(\theta,\phi)\delta(\theta',\phi')\rangle =
\sum_\ell C_\ell \left({2\ell+1\over 4\pi}\right) P_\ell(x)
\ee
where $x=\cos\psi$ and $P_\ell$ is a Legendre polynomial.  The
remainder of the calculation of the peak-peak correlation function
follows the method outlined in \scite{HS99}.  We compute the $12
\times 12$ covariance matrix $M_{ij} = \langle v_i v_j \rangle$, where
$v_i = (\bv_1, \bv_2)$ and the vectors $\bv$ specify the field and its
derivatives at the two points: $\bv = (\delta,\delta_\phi,\delta_\theta,
\delta_{\phi\phi},\delta_{\phi\theta},
\delta_{\theta\theta})$.  Note that $\delta_\phi \equiv
\partial\delta/\partial \phi$ etc.  
We show how to compute the components of the covariance matrix by an
example, from which the others can be readily generalised.  Consider
the correlation of derivatives in the $\phi$ direction at two points
(1) and (2):
\be
\langle \delta^{(1)}_\phi \delta^{(2)*}_\phi \rangle  =
\sum_{\ell,m}\sum_{\ell',m'} \langle a_{\ell m} a^*_{\ell'm'} \rangle
{\partial \over \partial \phi_1}Y_\ell^m(\theta_1,\phi_1)
{\partial \over \partial \phi_2}Y_{\ell'}^{m'*}(\theta_2,\phi_2)
\ee
We take the derivatives outside the summation, use the orthogonality
of the $a_{\ell m}$ (\ref{Cl}), and use the addition theorem for
spherical harmonics:
\be
\sum_m  Y_\ell^m(\theta_1,\phi_1)
Y_{\ell}^{m*}(\theta_2,\phi_2)={2\ell+1\over 4\pi} P_\ell(x).
\ee
This yields
\be
\langle \delta^{(1)}_\phi \delta^{(2)*}_\phi \rangle  =
{\partial^2 \over \partial \phi_1 \partial \phi_2}\sum_{\ell}\,
{2\ell+1\over 4\pi} C_\ell P_\ell(x).
\ee
Writing
$x=\cos\theta_1\cos\theta_2+\sin\theta_1\sin\theta_2\cos(\phi_1-\phi_2)$
we can differentiate to compute the covariance matrix element.  This
is aided by noting that these functions are independent of the
absolute positions or orientations of the two points on the sphere,
depending only on their separation.  We can therefore simplify the
algebra by taking $\theta_1=\theta_2=\pi/2$, $\phi_1=0$, and
$\phi_2=\phi$.  This element simplifies to
\be
\langle \delta^{(1)}_\phi \delta^{(2)*}_\phi \rangle  = \sum_\ell
{2\ell+1\over 4\pi} C_\ell \left[{dP_\ell(x)\over
dx}\cos\phi-{d^2P_\ell(x)
\over dx^2}\sin^2\phi\right].
\ee
Other elements are readily obtained by similar methods using Mathematica.

We invert $M$ to get the joint probability
distribution for the 12 variables,
\be
p(\bv_1,\bv_2) = {1\over (2\pi)^6 ||M||^{1/2}}
\exp\left(-{1\over 2} v_i M^{-1}_{ij} v_j\right).
\ee
and integrate subject to constraints that the two points are maxima:
\ba
1+\xi(r| \nu_1, \nu_2) &= & {1\over 4\theta_*^4  n_{pk}(\nu_1) n_{pk}(\nu_2)}
\int_{X_1=0}^\infty\int_{X_2=0}^\infty
\int_{Y_1=-X_1}^{X_1}\int_{Y_2=-X_2}^{X_2}
\int_{Z_1=-\sqrt{X_1^2-Y_1^2}}^{\sqrt{X_1^2-Y_1^2}}
\int_{Z_2=-\sqrt{X_2^2-Y_2^2}}^{\sqrt{X_2^2-Y_2^2}}
dX_1 dX_2 dY_1 dY_2 dZ_1 dZ_2\nn
&\times & \left(X_1^2-Y_1^2-Z_1^2\right)\left(X_2^2-Y_2^2-Z_2^2\right)
p(\nu_1,X_1,Y_1,Z_1,\eta^{(1)}_{\phi, \theta}=0,
\nu_2,X_2,Y_2,Z_2,\eta^{(2)}_{\phi,\theta}=0).
\label{Xsi}
\ea
where $n_{pk}(\nu)d\nu$ is the number density of peaks between height
$\nu$ and $\nu + d\nu$, given by A1.9 of \scite{BE87}.  By symmetry,
(\ref{Xsi}) is also the correlation function of minima at $-\nu_1$,
$-\nu_2$.  We have defined the symbols
\ba
\nu & \equiv & {\delta\over \sigma_0}\nn
\eta_{\phi} & \equiv & {\delta_{\phi}\over \sigma_1}\nn
\eta_{\theta} & \equiv & {\delta_{\theta}\over \sigma_1}\nn
X & \equiv & -{(\delta_{\phi\phi}+\delta_{\theta\theta})\over \sigma_2}\nn
Y &  \equiv & {(\delta_{\phi\phi}-\delta_{\theta\theta})\over \sigma_2}\nn
Z & \equiv & {2\delta_{\phi\theta}\over \sigma_2}
\label{Vars}
\ea
and the moments of the power spectrum are defined by
\ba
\sigma_0^2 &\equiv &F(1)\nn
\sigma_1^2 &\equiv &2F'(1)\nn
\sigma_2^2 &\equiv &4\left[F'(1)+2F''(1)\right]
\ea
where $F'(1)=dF(x)/dx|_{x=1}$ etc. We also define the spectral parameters
\be
\gamma \equiv \sigma_1^2/(\sigma_0
\sigma_2)  \qquad\theta_* \equiv \sqrt{2}{\sigma_1\over \sigma_2}.
\label{Parameters}
\ee
These allow simplification of the covariance matrix, with variables in
the order  ($\nu_1,\eta_{\phi 1},X_1,Y_1,\nu_2,\eta_{\phi
2},X_2,Y_2,\eta_{\theta 1},Z_1,
\eta_{\theta 2},Z_2$), to the block form
\be
M_{ij} = \left(
\begin{array}{ccc}
A & B & 0 \\
B^T & A & 0 \\
0 & 0 & C \\
\end{array}
\right)
\ee
where
\be
A = \left(
\begin{array}{cccc}
1 & \gamma & 0 & 0 \\
\gamma & 1 & 0 & 0 \\
0 & 0 & {1\over 2} & 0 \\
0 & 0 & 0 & (1 - \theta_*^2)/2\\
\end{array}
\right).
\ee
Defining $h(x) \equiv F(x)/\sigma_0^2$, $S\equiv \sin\phi$ and
$C\equiv \cos\phi$,
\be
B=\left(
\begin{array}{cccc}
h &
{\frac{{\theta_*^2}\, \left(2\,C\,h'-{S^2}\,h'' \right)}{2\,\gamma }}&
-{\frac{S\,\theta_* \,h'}{{\sqrt{2}}\,\gamma }} &
{\frac{{S^2}\,{\theta_*^2}\,h''}{2\,\gamma }}\cr
B_{21} &
{\frac{{\theta_*^4}\,\left[ 4\,C\,h'+\left( 8\,{C^2}-6\,{S^2}\right)\,h''-
8\,C\,{S^2}\,h^{(3)} + {S^4}\,h^{(4)} \right] }{4\,{{\gamma }^2}}} &
{\frac{S\,{\theta_*^3}\,\left( -2\,h' - 4\,C\,h'' + {S^2}\,h^{(3)}
\right) }{2\,{\sqrt{2}}\,{{\gamma }^2}}} & {\frac{{\theta_*^4}\,
\left( 6\,{S^2} \,h'' +
6\,C\,{S^2}\,h^{(3)} - {S^4}\,h^{(4)} \right) }{4\,{{\gamma }^2}}}\cr
-B_{13} & -B_{23} &
{\frac{{\theta_*^2}\,\left( C\,h' - {S^2}\,h'' \right) }
{2\,{{\gamma }^2}}} &
{\frac{S\,{\theta_*^3}\,\left( -2\,C\,h'' + {S^2}\,h^{(3)} \right) }{2\,
{\sqrt{2}}\,{{\gamma }^2}}}
\cr
B_{14} & B_{24} & -B_{34} &
      {\frac{{\theta_*^4}\,
      \left[ 2\,\left( 1 + {C^2} \right) \,h'' - 4\,C\,{S^2}\,h^{(3)} +
      {S^4}\,h^{(4)} \right] }{4\,{{\gamma }^2}}} \cr
\end{array}
\right)
\ee
where $h^{(3)}(x) \equiv h'''(x)$ etc, and we write the lower triangle
in terms of the upper triangular matrix for conciseness.  Finally,
\be
C=\left(\begin{array}{cccc}
{\frac{1}
    {2}} & {\frac{{\theta_*^2}\,h'}{2\,{{\gamma }^2}}} & 0 & -{\frac{S\,
       {\theta_*^3}\,h''}{{\sqrt{2}}\,{{\gamma }^2}}} \cr
{\frac{{\theta_*^2}\,h'}{2\,{{\gamma }^2}}} & {\frac{1}{2}} & {
     \frac{S\,{\theta_*^3}\,h''}{{\sqrt{2}}\,{{\gamma }^2}}} & 0
\cr
0 & {\frac{S\,{\theta_*^3}\,h''}
    {{\sqrt{2}}\,{{\gamma }^2}}} & {\frac{1 - {\theta_*^2}}{2}} & {\frac{
      {\theta_*^4}\,\left( C\,h'' - {S^2}\,h^{(3)} \right)
}{{{\gamma }^2}}}
    \cr -{\frac{S\,{\theta_*^3}\,h''}
     {{\sqrt{2}}\,{{\gamma }^2}}} & 0 & {\frac{{\theta_*^4}\,
      \left( C\,h'' - {S^2}\,h^{(3)} \right) }{{{\gamma }^2}}} & {\frac{1 -
      {\theta_*^2}}{2}} \cr
\end{array}\right)
\ee
The correlation function for peaks above a certain threshold $\nu$ is
obtained by adding two further integrations over $\nu_1$ and $\nu_2$,
and replacing the differential number densities $n_{pk}(\nu)$ in the
denominator of (\ref{Xsi}) by numerically-evaluated integrals
$n_{pk}(>\nu)$.  For peaks above a threshold, the 8D integration can
be reduced to 6, as the integrals over $\nu_2$ and $z_2$ can be done
analytically.  Very accurate integrations can then be done on a desktop
workstation in about 50 seconds.

\section{Results}

\begin{figure}
\centering
\begin{picture}(200,200)
\includegraphics{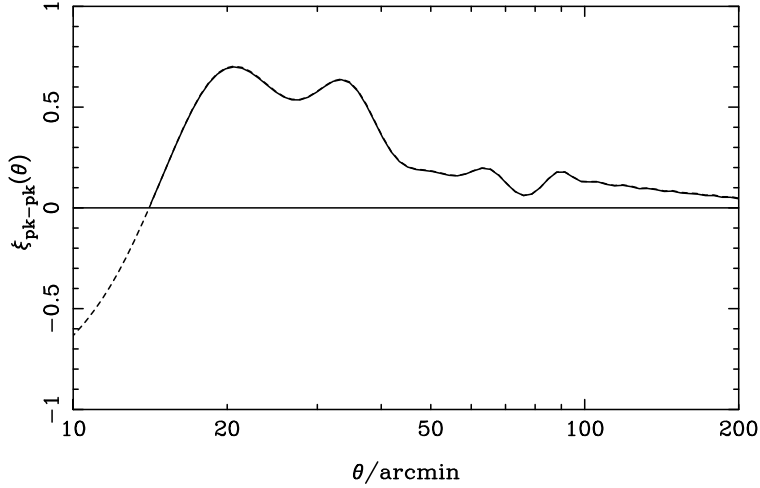}
\end{picture}
\caption{\label{XsiPlot} (Solid line) The correlation function for
peaks above a +1$\sigma$ threshold, in a mixed dark matter model with
CDM, vacuum and baryon density parameters $\Omega_{CDM}=0.8$,
$\Omega_\nu=0.15$ and $\Omega_B=0.05$. Hubble constant is $H_0=60$ km s$^{-1}$
Mpc$^{-1}$.  For comparison, the flat sky results of Heavens \& Sheth
(1999) are shown dotted.  The results coincide to an accuracy of
better than 0.004.}
\end{figure}

\begin{figure}
\centering
\begin{picture}(200,200)
\includegraphics{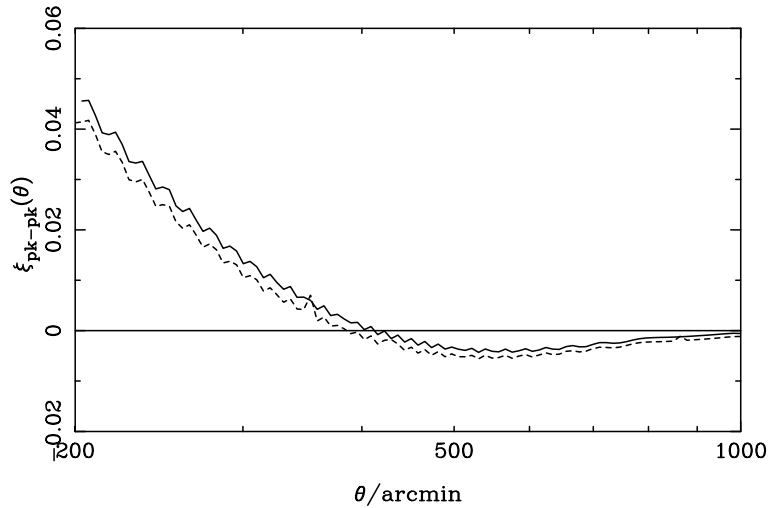}
\end{picture}
\caption{\label{XsiPlotLA} As Fig. \ref{XsiPlot}, but at larger angle
separations between 3.3 and 16.7 degrees.}
\end{figure}

We run CMBFAST \cite{SZ96} to generate the power spectrum $C_\ell$,
and model the beam with a gaussian of FWHM $b$, so multiply the power
spectrum by a gaussian $\exp\left[-\sigma^2 \ell(\ell+1)\right]$, with
$\sigma= b/\sqrt{8\ln2}$.  
We have not included the effects of gravitational lensing on the
temperature field.  As shown by \scite{Takada2000}, the effect is small
except for separations up to the first peak, where the anticorrelation
is reduced in magnitude.  Figs. \ref{XsiPlot} and \ref{XsiPlotLA} show 
the correlation function of peaks above a $1\sigma$ threshold for a
mixed dark matter model, along with the results of the flat-sky
calculation of \scite{HS99}.  The differences are at the level of
$\sim 0.005$.


\section{Correlation function vs bispectrum for string maps}

\begin{figure}
\centering
\begin{picture}(200,200)
\includegraphics{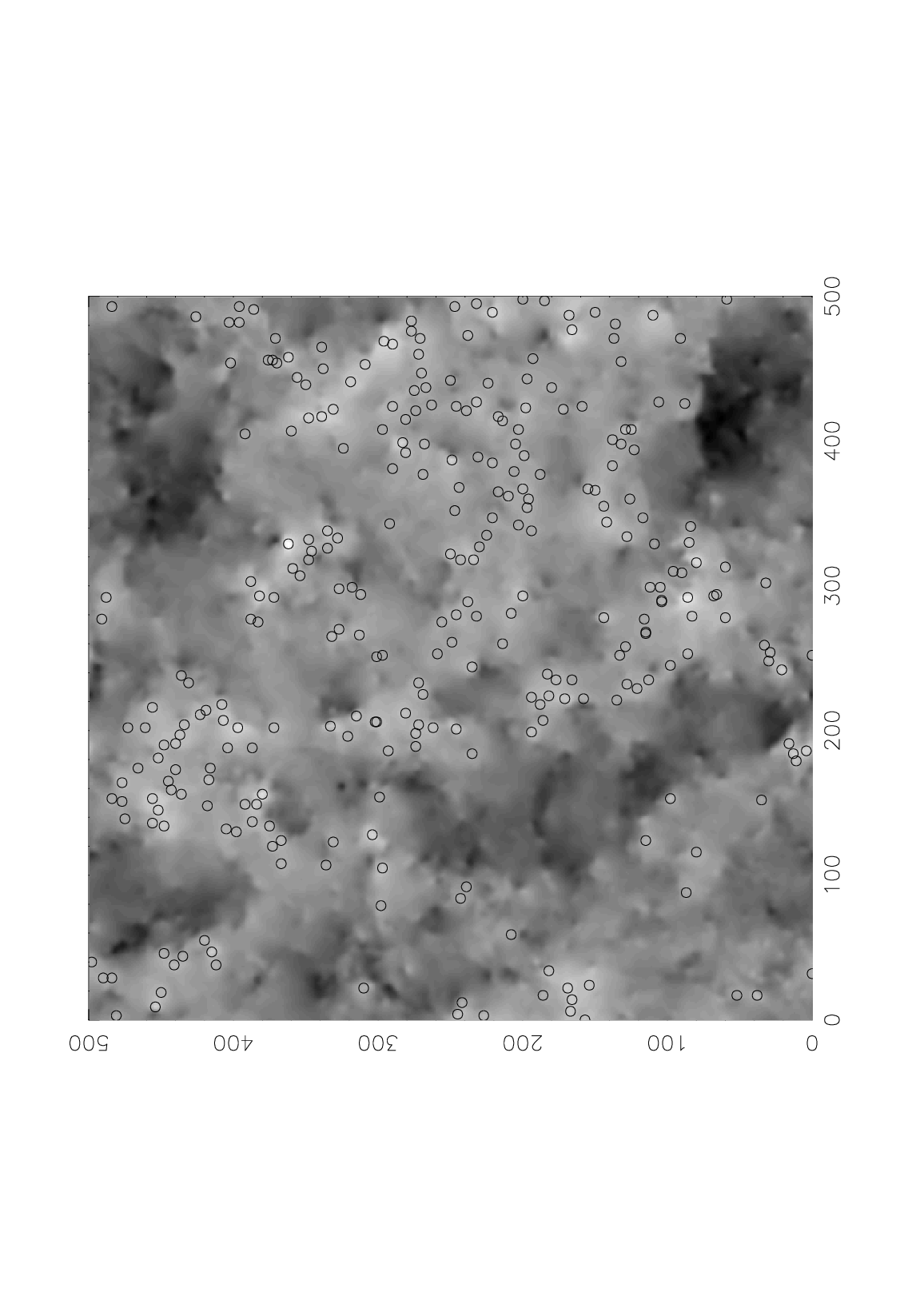}
\end{picture}
\caption[]{\label{StringMap} Simulated sky map for cosmic strings, 
consisting of a gaussian background, with the lensing effect of a
string network superimposed.  Peaks above $1\sigma$ are circled.}
\end{figure}

There are many methods for testing the gaussian hypothesis, and
it is tempting to ask which is the best.  Unfortunately the
question is badly posed, as methods will fare differently
depending on the exact properties of the non-gaussian field
considered.  Here we focus on one particular non-gaussian field,
produced by a network of cosmic strings. Fig. \ref{StringMap}
shows a realisation of the temperature map expected from cosmic
strings, one of two kindly provided by Francois Bouchet.  The
lensing effect of the moving strings is added to a gaussian
background map, approximately as expected from the string model
\cite{PST97} (see also \pcite{SP2000}, \pcite{Avelino2000}).  
We consider two diagnostics: the peak-peak
correlation function, and the bispectrum (e.g. \pcite{Hea98},
\pcite{Ferreira98}, \pcite{GM2000}).  The string maps we have are 12.5$^\circ$ on a
side, and we would expect the bispectrum to have difficulty in
distinguishing these maps from gaussian maps with the same power
spectrum \cite{Luo94}.  The interesting question is whether the
peak correlation function can do better.
Since the string simulations are performed on a small, flat patch
of sky, we use a Fourier transform, to compute the flat-sky
bispectrum from the Fourier coefficients $\delta(\bf k)\equiv \int
d^2{\bf x} \delta({\bf x})\exp(i\bkk\cdot {\bf x})$:
\be 
\langle \delta (\bkk{1}) \delta (\bkk{2})
\delta(\bkk{3})\rangle =(2\pi)^{2}B(\bkk{1},\bkk{2},\bkk{3})
\delta^{D}(\bkk{1}+\bkk{2}+\bkk{3}) 
\ee
The angle brackets indicate ensemble averages, and $\delta^D$ is the
Dirac delta function. For a gaussian field the bispectrum is zero, and
for all fields the bispectrum is zero unless
$\bkk{1}+\bkk{2}+\bkk{3}={\bf 0}$. Following the work of \scite{MVH97}
and \scite{VHMM98} in large-scale structure, we consider two
configurations of triangles: equilaterals, and zero-area triangles
with two equal wavevectors and one of twice the size.  The bispectrum
is real, but the $\delta({\bf k})$ are not, so we consider the estimate
\be D_{\alpha}=\langle Re(\delta({\bf k}_{1})\delta({\bf
k}_{2})\delta({\bf k}_{3}))\rangle \ee
and average over thin shells in $k-$space.

\begin{figure}
\centering
\begin{picture}(200,200)
\includegraphics{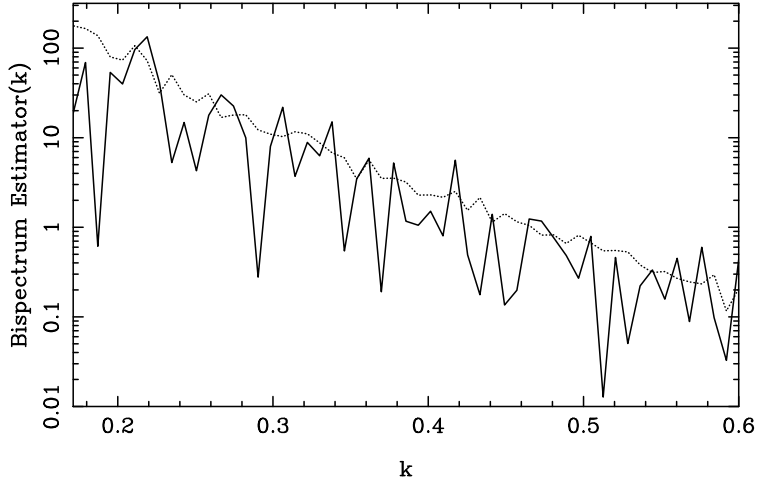}
\end{picture}
\caption{\label{Bequ} The equilateral bispectrum as estimated from 
map with string foreground and gaussian temperature on last-scattering 
surface (solid), and cosmic r.m.s. (dotted).}
\end{figure}

Fig. \ref{Bequ} displays the equilateral estimated
bispectrum for the string map shown in Fig. \ref{StringMap}. Also
shown is the cosmic r.m.s. for a gaussian field of the same power
spectrum, $\langle |\delta_{\bf k}|^2\rangle^{3/2}$.
We show in  Table \ref{tablewigauss} reduced $\chi^2$ values for
both equilateral triangles and zero-area triangles for the two
simulated maps. With 55 bins, the variance in the reduced $\chi^2$
for a gaussian field is shown in the final column.  We find no
significant departure from gaussianity with this test.
\begin{table}
\centering
\begin{tabular}{||c||c||c|c||}           \hline
 &equilateral&zero-area&$\sqrt{2/(n-1)}$     \\ \hline
Map 1&1.06 &0.96     & 0.19            \\ \hline
Map 2&1.29 &1.42     & 0.19            \\ \hline
\end{tabular}
\caption[]{\label{tablewigauss} Reduced $\chi$-squared values of
the deviation of the bispectra from a gaussian model of the two
modified maps, consisting of intrinsic gaussian and cosmic string
generated fluctuations.  The r.m.s. of the reduced $\chi^2$ for a
gaussian model with $n=55$ data is shown in the final column.}
\end{table}
\begin{figure}
\centering
\begin{picture}(200,200)
\includegraphics{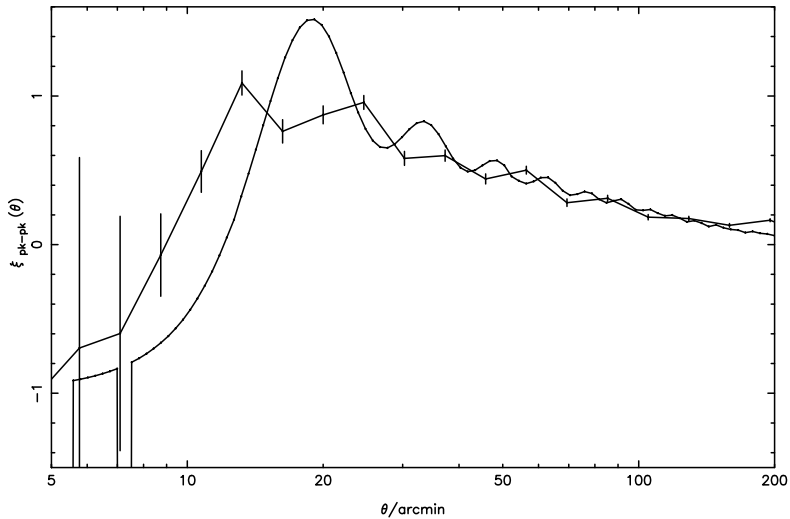}
\end{picture}
\caption{\label{pk_pk_mp1_stringsonly} The correlation function of
peaks above $1\sigma$ calculated from the map of Fig. \ref{StringMap}.
Errors are Poisson, and hence underestimates.  Superimposed is the
correlation function from a gaussian map with the same power spectrum.
Note the excess of string peaks around 10-15 arcminutes.}
\end{figure}

Fig. \ref{pk_pk_mp1_stringsonly} shows the correlation function
of peaks above $1\sigma$ (where $\sigma^2$ is the map variance)
for the map shown in Fig. \ref{StringMap}.  The map is smoothed
with a gaussian beam of FWHM 5.5$'$ to model the Planck beam.
The errors for the peak-peak correlation function are Poisson
errors, which will be underestimates.  However, it is clear that
the peak correlation function of the string map is significantly
different from that of a gaussian map with the same power
spectrum.  The most striking difference is the presence of peaks
in the string map which are separated by $10-20$ arcminutes.
These appear in greater numbers than in the gaussian map, and
this could be the most obvious manifestation of strings.

\section{Discussion}

We have presented calculations of the exact correlation function
of peaks in a random gaussian field defined on the surface of a
sphere.  No small-angle approximation is made, so the method is
an advance on the flat-sky computations of \scite{HS99} and now
effectively complete. The formalism allows very accurate
theoretical predictions of the peak-peak correlation function for
temperature fluctuations in the microwave background, which is the
application considered here. We envisage the main use of this
method being as a sensitive test of the gaussian hypothesis.
Since inflationary models generically predict a temperature field
which is very close to gaussian, this is a consistency test for
inflation. Other structure formation models, based for example on
strings, predict non-gaussian temperature maps.  Although the
visual appearance of string maps is evidently non-gaussian, it is
not necessarily easy to find statistics which will unambiguously
distinguish them from gaussian fields.  To illustrate this point,
we have analysed 12.5-degree square simulated maps of string
models, using the bispectrum and the peak-peak correlation
function as distinguishing statistics.  We find that, while
cosmic variance in the bispectrum makes it difficult to use on a
small patch of sky, the peak-peak correlation function clearly
rules out a gaussian map.

In practice, maps of the microwave background will be
contaminated at some level by point sources, amongst other
things.  Peak statistics may be useful in assessing this
contribution.  The most straightforward example is that an
uncontaminated map has the same average number density of maxima
and minima; a significant excess of maxima would be indicative of
contamination.  Unfortunately the theory of peaks is not able to
tell us the {\em distribution} of the number of maxima or minima
within a finite sky (only its mean), but it is a straightforward
matter to determine the distribution by monte carlo
realisations.  One can attempt to go further than this, by
removing statistically the contribution from the point sources,
provided one knows from other observations what their correlation
function is. Assuming the point sources are uncorrelated with the
microwave background peaks, the correlation function of the
combined map is simply a weighted mean of the two.  The point
sources will contribute to the power spectrum; one can vary the
assumed contribution from point sources and modify the power
spectrum and the derived microwave background peak correlation
function accordingly.  If consistency can be achieved, one will
be confident both of the gaussian nature of the microwave
background, and the level of point source contamination.

{\bf Acknowledgments}

We are grateful to Francois Bouchet for providing the simulated
catalogues, and to Ravi Sheth for useful discussions.
Computations were made partly using Starlink facilities.



\end{document}